\documentstyle[aps,prb,epsf,floats]{revtex}

\begin{document}
\twocolumn[\hsize\textwidth\columnwidth\hsize\csname@twocolumnfalse%
\endcsname

\title {Quantum phase transitions in models of coupled magnetic impurities}

\author{Matthias Vojta, Ralf Bulla, and Walter Hofstetter}
\address{Theoretische Physik III, Elektronische Korrelationen und
Magnetismus, Universit\"at Augsburg, 86135 Augsburg, Germany}
\date{June 15, 2001}
\maketitle

\begin{abstract}

We discuss models of interacting magnetic impurities coupled to
a metallic host,
which show one or more boundary quantum phase transitions
where the ground-state spin changes as a function of the
inter-impurity couplings.
The simplest example is realized by two spin-1/2 Kondo impurities
coupled to a single orbital of the host.
We investigate the phase diagram and crossover behavior
of this model and present Numerical Renormalization Group
results together with general arguments
showing that the singlet-doublet quantum phase transition
is either of first order or of the Kosterlitz-Thouless
type, depending on the symmetry of the Kondo couplings.
Thus we find an exponentially small energy scale within
the Kondo regime, and a two-stage Kondo effect, in a
single-channel situation.
Connections to other models and possible applications
are discussed.

\end{abstract}
\pacs{PACS numbers:}
\vspace*{-15 pt}
]


The physics of a single magnetic impurity embedded in a metal --
known as the Kondo effect -- is a well studied phenomenon in
many-body physics~\cite{hewson}.
The low-energy physics 
is completely determined by a single energy scale,
the Kondo temperature $T_{\rm K}$,
and the impurity spin is fully quenched in the low-temperature limit,
$T \ll T_{\rm K}$.

Systems with more than one impurity are considerably more complicated
because the interaction between impurity spins competes with 
Kondo screening.
This inter-impurity interaction, which can lead to a magnetic ordering
transition in lattice models, arises both from direct exchange
coupling $I$, which we explicitely include in the following
discussion, and from the Ruderman-Kittel-Kasuya-Yosida (RKKY)
interaction mediated by the conduction electrons.

The simplest systems containing this competition between Kondo effect
and inter-impurity interaction are two-impurity models
\cite{2imp,2impnrg,2impsakai,2impcft,2impoli}
which have been extensively studied in recent years.
Generically, two different regimes are possible as function
of the inter-impurity exchange $I$:
for large antiferromagnetic $I$ the impurities combine to
a singlet, and the interaction with the conduction
band is weak, whereas for ferromagnetic $I$ the impurity
spins add up and are Kondo-screened by conduction
electrons in the low-temperature limit.

For two spin-1/2 impurities, each coupled to one channel of
conduction electrons (total number of two screening channels $K\!=\!2$),
there is perfect screening as $T \to 0$ in both regimes and the
ground-state spin will be zero.
It has been shown that there is
{\em no} quantum phase transition as $I$ is varied
in this two-impurity Kondo model
in the generic situation without particle-hole symmetry
(whereas one finds an unstable non-Fermi liquid fixed
point in the particle-hole symmetric
case)~\cite{2impnrg,2impsakai,2impcft}.
However, the finite-$T$ crossover behavior is significantly
different in the two regimes:
for large antiferromagnetic $I$, the singlet is
formed at an energy scale of order $I$.
In contrast, for ferromagnetic $I$ where both
spins have to be quenched by the Kondo effect, the two screening
channels are formed by the even and odd linear combinations
of the conduction electrons at both impurity sites~\cite{2imp}.
These two channels have generically different
densities of states leading to two different Kondo
temperatures.
Upon lowering the temperature the two spin-1/2 degrees of freedom
are ``frozen'' at these two distinct temperature scales
(two-stage Kondo effect \cite{2imp,2impoli}).

The purpose of this paper is to discuss models of coupled
impurities which can be tuned between regimes with
{\em different} ground state spin.
The spin quantization then {\em requires} that these regimes
are separated by one or more boundary quantum phase transitions
as the inter-impurity couplings are varied.
Phases with a non-zero ground state spin are
obtained in the situation of underscreening \cite{NB},
{\em i.e.}, if a spin of size $S$ interacts with $K$
conduction electron channels, and $2S > K$, leading
to a residual free spin of size $(S\!-\!K/2)$ as $T\to 0$.
A general 
two-impurity Kondo model
necessarily showing a quantum phase transition
then consists of
two spins of size $S_1$ and $S_2$, with a direct
exchange coupling $I$, and a total of $K$  
screening channels,
satisfying $2(S_1\!+\!S_2) > K$.

Related Kondo- and Anderson models have been
discussed recently in the context of multi-level
or coupled quantum dots with nearly
degenerate singlet and triplet levels
\cite{qdotexp,qdotth0,qdotth1,qdotth2,qdotth4}.
Several authors have considered effective two-impurity models
near the singlet-triplet degeneracy
with {\em two} screening channels \cite{qdotth2} which
arise in vertical multi-level dots;
such a situation corresponds to the physics of the usual
two-impurity Kondo model {\em not} showing a phase transition
in the absence of particle-hole symmetry \cite{2impcft}.
The case of a single screening channel and the associated
limiting situations of a singlet ground state and
an underscreened Kondo model have been mentioned in
Refs.~\cite{qdotth1,qdotth4},
but the nature of the transition and the associated
quantum-critical behavior have not been examined
to date.

We note that a two-impurity model with $2(S_1\!+\!S_2) < K$
can give rise to overscreening \cite{NB} in a channel-symmetric
situation.
Overscreening leads to a non-trivial intermediate coupling
fixed point already for a single impurity and is a very
interesting topic on its own, which shall, however, not be
discussed here.
Remarkably, 
the two-impurity two-channel Kondo model ($K\!=\!4$),
has been solved exactly by Georges and Sengupta, and
shows a continuous line of non-Fermi liquid fixed
points \cite{2imp2ck}.


\begin{figure}[t]
\epsfxsize=3in
\centerline{\epsffile{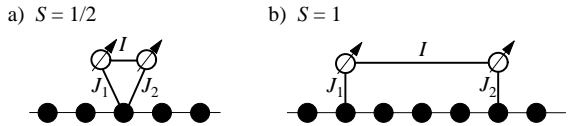}}
\caption{
Kondo models discussed in the text.
The solid/empty dots represent conduction electron orbitals
and Kondo impurities, respectively.
$J_{1,2}$ denote the Kondo couplings,
$I$ is an explicit inter-impurity exchange.
a) Two spins 1/2, one channel.
b) Two spins 1, two channels \protect\cite{lehur}.
}
\vspace{-8pt}
\label{figmodel}
\end{figure}

{\it Two spins 1/2, one channel}.
The simplest realization of the above proposal is a
``double Kondo impurity'' (Fig.~\ref{figmodel}a),
{\em i.e.}, a Kondo model
of two spins 1/2 coupled to a single orbital of conduction
electrons, with the Hamiltonian $H=H_{\rm band}+H_{\rm int}$,
\begin{equation}
H_{\rm int} =
    J_1 {\bf S}_1 \cdot {\bf s}_0
  + J_2 {\bf S}_2 \cdot {\bf s}_0
  + I {\bf S}_1   \cdot {\bf S}_2
\label{H0}
\end{equation}
and
$H_{\rm band} = \sum_{{\bf k}\alpha} \epsilon_{\bf k} c^\dagger_{{\bf k}\alpha} c_{{\bf k}\alpha}$
in standard notation,
${\bf s}_0 = \sum_{\bf kk'\alpha\beta} c^\dagger_{{\bf k}\alpha} {\bf \sigma}_{\alpha\beta}
c_{{\bf k '}\beta}$ is the conduction band spin operator at the impurity
site ${\bf r}_0 = 0$, and ${\bf S}_{1,2}$ denote the impurity spin operators.
The Kondo couplings $J_1$, $J_2$ are taken to be antiferromagnetic.
For the numerical calculations, we will use a conduction band with
a constant density of states,
$\rho(\epsilon) = \rho_0 = 1/(2D)$ for $|\epsilon| < D = 1$.

The inter-impurity coupling $I$ obviously drives a quantum phase
transition:
For large antiferromagnetic $I$, both impurities form a singlet
with zero total spin, whereas large ferromagnetic $I$ locks the
impurities into an $S=1$ entity, which is underscreened by the
single channel of conduction electrons, leading to
a doublet ground state with residual spin 1/2.

A second observation is that the symmetry w.r.t. to the two
Kondo couplings plays a crucial role.
Re-writing $H_{\rm int}$ (\ref{H0}) in terms of bond
boson operators \{$s,t_{x,y,z}$\} \cite{sachdev_1},
with
$S_{1,2}^{\alpha} = ( \pm s^{\dagger}  t_{\alpha}^{} \pm
t_{\alpha}^{\dagger} s^{}  - i \epsilon_{\alpha\beta\gamma} t_{\beta}^{\dagger}
t_{\gamma}) / 2$,
one finds
\begin{eqnarray}
H_{\rm int} =
 I \, ( t_{\alpha}^\dagger t_{\alpha} - 3 s^\dagger s)/4
&-& (J_1 + J_2) \,  i \epsilon_{\alpha\beta\gamma} {\bf s}_0^\alpha t_{\beta}^{\dagger} t_{\gamma}
\nonumber \\
&+& (J_1 - J_2) \, {\bf s}_0^\alpha (s t_\alpha^\dagger + h.c.) .
\nonumber
\end{eqnarray}
In the symmetric case, $J_1=J_2$, there is no mixing
between impurity singlet and triplets,
and the state formed by the inter-impurity singlet and a decoupled
conduction band is an exact eigenstate of $H$.
The ``rest'' of the Hilbert space (triplet plus conduction band)
forms an underscreened $S=1$ Kondo problem.
Therefore, for $J_1=J_2$ the model shows a simple level crossing at a
critical $I_c$ which marks a first-order transition between
the singlet and the underscreened doublet state.

\begin{figure}[t]
\epsfxsize=3.2in
\centerline{\epsffile{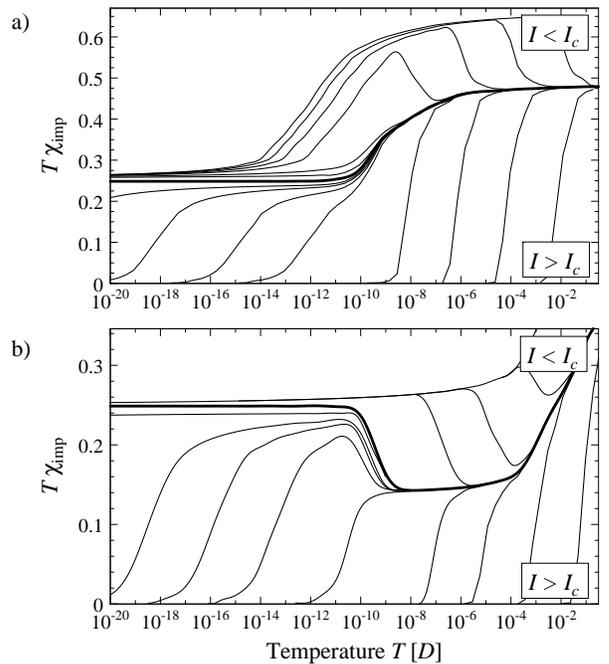}}
\caption{
NRG results \protect\cite{nrgprob}
for $T\chi_{\rm imp}$ ($k_B\!=\!1$)
of the double-impurity
model.
a) Large asymmetry: $J_1 \!=\! 0.1$, $J_2 \!=\! 0.05$.
The thick curve corresponds to $I \!=\! 0.00181527408 \!\simeq\! I_c$,
the curves close to $I_c$ are for $10^{11} (I-I_c) = -3, +2, +3, +5$,
the others for $|I-I_c| = 10^{-10}, 10^{-8}, ..., 10^{-2}$.
b) Small asymmetry: $J_1 \!=\! 0.4$, $J_2 \!=\! 0.400001$.
Here $I_c \simeq 0.0788628547$,
curves are displayed for $10^{10} (I-I_c) = +0.5, +1, +1.2, +1.5, +7.5$
and $|I-I_c| = 10^{-7}, ..., 10^{-1}$.
}
\vspace{-8pt}
\label{fignrg1}
\end{figure}

Turning to the asymmetric case, $J_1>J_2$, we start with
results obtained using the Numerical
Renormalization Group (NRG) technique \cite{nrg}.
In Fig.~\ref{fignrg1} we display the temperature dependence of
$T\chi_{\rm imp}$ for fixed $J_{1,2}$ and different values of $I$,
where $\chi_{\rm imp}$ is the total impurity
contribution to the uniform susceptibility.
Let us first discuss the case of large asymmetry, Fig.~\ref{fignrg1}a.
We can identify $I_c$, with all curves for $I>I_c$ leading to
$T\chi_{\rm imp}\to 0$ for $T\to 0$ (singlet),
whereas all curves for $I\leq I_c$ show $T\chi_{\rm imp}\to 1/4$,
consistent with a residual spin 1/2.
The topmost curves ($I \ll I_c$) allow an identification of
the Kondo temperature for the $S=1$ Kondo problem,
here we have $T_K \simeq 10^{-11} D$
(small values of $J_{1,2}$ have been chosen to illustrate the crossover
behavior).
The curves for $I\! <\! I_c$ approach the fixed point very slowly,
indicating an operator which is marginally irrelevant.
For $I$ values near $I_c$ and $T \gg T_K$, there is a plateau
at $T\chi_{\rm imp} \simeq 1/2$.
For exchange couplings $I$ larger than but close to $I_c$, the
system first flows to the fixed point with $T\chi_{\rm imp} = 1/4$,
and eventually crosses over to the singlet ground state
(for $T/D < 10^{-12}$ in  Fig.~\ref{fignrg1}a).
The corresponding crossover scale $T^\ast$ is found to
depend {\em exponentially} on the distance to the critical point,
$(I-I_c)$ -- there is no power law which would characterize
a second-order transition.
Even for very small $|I-I_c|$ the NRG does not show any
fixed point behavior which could be distinguished from one of the
stable fixed points, {\em i.e.}, there is no sign of a third,
infrared unstable, fixed point which could possibly correspond
to the critical point at $I=I_c$.

We have analyzed the structure of the NRG fixed points,
and found that for $I>I_c$ the system consists of a singlet
decoupled from the band ($\langle {\bf S}_1 \cdot {\bf S}_2\rangle = -3/4$),
whereas for all other values
{\em including} $I_c$ there is one Kondo-screened
spin 1/2 and one residual decoupled spin 1/2.
Close to $I=I_c$, the NRG flow is identical
to the one in the single-impurity Kondo model
near the Kosterlitz-Thouless (KT) transition at $J=0$.
These results strongly suggest that the phase transition
in the double-impurity model
is of KT type, and the critical point
is characterized by one decoupled spin-1/2 degree
of freedom.

In the following, we give heuristic scaling arguments
supporting this picture.
Close to the critical point, we can assume that the
effective inter-impurity coupling, $I_{\rm eff} = I + I_{\rm RKKY}$,
is the smallest energy scale of the problem.
In the spirit of poor man's scaling, both Kondo couplings will grow
upon reducing the band cut-off, and the initial asymmetry between
$J_1$ and $J_2$ increases under this process.
The larger coupling, $J_1$, will become of order unity at an
energy scale $T_K \sim D \exp[-D/(2J_1)]$,
indicating that the spin ${\bf S}_1$ becomes screened.
This screening freezes the spin ${\bf S}_1$ and the conduction
electron spin ${\bf s}_0$ into a singlet,
thus terminating the flow of $J_2$ and effectively decoupling
the spin ${\bf S}_2$ from the band.

If we now switch on a small $I_{\rm eff} \ll T_K$, it will
couple the remaining spin {\em via the screened singlet}
to the conduction band.
This effective Kondo coupling can be estimated by second-order
perturbation theory to be of order $I_{\rm eff} D / T_K$.
Therefore, ferromagnetic $I_{\rm eff}$ flows to zero
as known from the ferromagnetic Kondo effect, and
one ends up in a weak-coupling regime for the second
spin, with a doublet ground state.
On the other hand, antiferromagnetic $I_{\rm eff}$
grows logarithmically and leads to screening
of the second spin 1/2 \cite{KondoKT}.
But, in contrast to a single-spin Kondo effect, this flow is
cut-off when the effective coupling of the second spin
becomes comparable to the singlet binding energy
of the first spin ${\bf S}_1$, $T_K$, due to the indirect nature
of this exchange.
It is immediately clear what happens at this scale:
The two spins ${\bf S}_1$ and ${\bf S}_2$
acquire a large direct antiferromagnetic
coupling, {\em i.e.}, they form a singlet among themselves
and effectively decouple from the band.
Therefore, this regime of weak antiferromagnetic
$I_{\rm eff}$ is continuously connected to the one
of strong antiferromagnetic $I$ where the two spins form
an inter-impurity singlet from the outset.
We can identify $I_{\rm eff}$ with $(I-I_c)$,
and the crossover scale $T^\ast$ near the critical point is given
by $T^\ast \sim D \exp(-T_K/[I-I_c])$.

In the case of maximal asymmetry, $J_2=0$,
the picture developed above is most transparent.
Here, no RKKY interaction is generated, and the critical
coupling value is obviously $I_c =0$.
For $J_2>0$, we find $I_c \sim J_1 J_2 / D > 0$ since the
RKKY coupling is generically ferromagnetic because
the two antiferromagnetic Kondo couplings prefer
a parallel alignment of ${\bf S}_1$ and ${\bf S}_2$.
Similarly, $J_2<0$ (but $J_1>0$) gives a KT
transition at $I_c < 0$. \cite{ferro}

\begin{figure}[t]
\epsfxsize=3.5in
\centerline{\epsffile{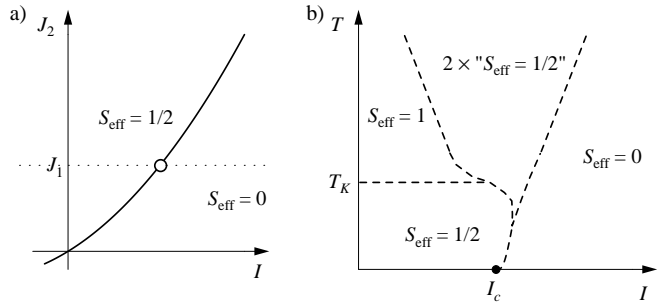}}
\caption{
Schematic phase diagrams of the double-impurity
Kondo model (Fig.~\protect\ref{figmodel}a).
The $S_{\rm eff}$ values in the different
regimes are defined through the Curie term in the
total impurity susceptibility,
$T \chi_{\rm imp} = S_{\rm eff}(S_{\rm eff}+1) / 3$.
a) For $T=0$ and fixed $J_1$.
The solid line is a line of KT transitions,
the open dot is the first-order transition for
$J_1=J_2$.
b) For $T\geq 0$ and fixed $J_{1,2}$ with
not too small asymmetry $(J_1\!-\!J_2)$.
The solid dot is the KT transition,
the dashed lines denote crossovers.
}
\vspace{-8pt}
\label{figpd1}
\end{figure}

Summarizing, the critical point in the double-impurity
problem is defined by the coupling value where
one spin completely decouples from the band, and the other
spin is Kondo-screened.
The boundary quantum phase transition is of the same type as in the
single-impurity Kondo model when the bare coupling is tuned
through zero, {\em i.e.}, of the KT
universality class \cite{KondoKT,book,KTremark}.
Close to the critical point for $I>I_c$, we
find a new exponentially small energy scale $T^\ast$
{\em within} the Kondo regime --
leading to two exponentially small scales, with $T^\ast \ll T_K$ --
and the crossovers can be understood as a {\em two-stage
Kondo effect}.
Such a two-stage behavior was so far only known in two-channel
models~\cite{2imp,2impoli};
remarkably, here it is realized with a single
conduction electron channel.

We emphasize, however, that the stable fixed point for $I>I_c$
is {\em not} given by two Kondo-screened impurities,
but by an inter-impurity singlet.
This has immediate consequences for the
conduction electron phase shift.
For ferromagnetic $I_{\rm eff}$ ($I < I_c$), there is one
screened spin 1/2 leading to a phase shift of $\pi/2$,
whereas
for $I > I_c$ the low-$T$ limit of the phase shift is zero.

It is also interesting to discuss the limit
of small asymmetry of the Kondo couplings.
As shown above, the singlet-doublet transition is
of first order in the symmetric case.
Therefore, very small asymmetry introduces
another low-energy scale, $T_{\rm as} \propto |J_1-J_2|$,
and the KT behavior will be visible only for
$T \ll T_{\rm as}$.
We have verified this behavior numerically, and sample results
are shown in Fig.~\ref{fignrg1}b.
Here, $T_{\rm as}\!\approx\! 10^{-10} D$, $T_K\!\approx\! 10^{-3} D$,
and a new regime
(compared to Fig.~\ref{fignrg1}a) occurs
for $T_{\rm as} < T < T_K$ where effectively level crossing
between the singlet and the underscreened doublet is observed.
Our numerical results are consistent with
$T_{\rm as} \sim T_K D |J_1-J_2|/(J_1 J_2) \sim |T_{K1} - T_{K2}|$,
with $T_{K1,2} \sim D \exp[-D/(2J_{1,2})]$.

Schematic phase diagrams inferred from both the numerical
results and the above arguments are shown in Fig.~\ref{figpd1}.
At $T=0$, Fig.~\ref{figpd1}a, we have a line of KT transitions
terminating at a first-order point for $J_1=J_2$.
The finite-$T$ crossovers are escpecially interesting, Fig.~\ref{figpd1}b:
Above $T_K$, the physics is dominated by the
crossing of the $S=0$ and $S=1$ impurity levels --
in particular, near $I=I_c$ the two spins 1/2 fluctuate
independently, leading to the plateau at
$T \chi_{\rm imp} \simeq 2 \times S(S\!+\!1) / 3$ with $S\!=\!1/2$.
In contrast, below $T_K$ the behavior is determined
by the KT quantum phase transition
at $I_c$.


{\it Two spins 1, two channels}.
We briefly want to comment on a different model,
first considered in Ref.~\onlinecite{lehur},
which consists of two spins 1 coupled to a total of
two screening channels ($K\!=\!2$),
{\em i.e.}, two underscreened $S=1$ Kondo impurities,
Fig.~\ref{figmodel}b.
The ground state spin can be tuned between zero and one,
and therefore there {\em must} be a quantum phase transition
as the coupling $I$ is varied.
Ref.~\onlinecite{lehur}
did not consider such a transition, but focussed exclusively on
a possible phase transition~\cite{2impnrg,2impsakai,2impcft}
associated with a critical non-Fermi liquid fixed point
and a jump in the phase shift.
It was found that such a transition does not exist
for spins 1.

In analogy to the discussion above, we expect that
at sufficiently low temperatures and $I_{\rm eff}\to 0$,
each impurity realizes an
underscreened $S=1$ Kondo effect, {\em i.e.}, a spin 1/2
remains unscreened at each of the impurity sites.
Turning on a ferromagnetic $I_{\rm eff}$ the two residual
spins couple to an unscreened spin 1.
In contrast, $I_{\rm eff} > 0$ promotes an effective
antiferromagnetic coupling, both among the two residual spins
as well as to the conduction electrons.
However, the energy gain due to singlet
formation with the conduction band is exponentially small
in $I_{\rm eff}$, whereas the energy of the inter-impurity singlet
is linear in $I_{\rm eff}$. Therefore, singlet formation
between the two residual spins 1/2 is favorable, and this
singlet will be effectively decoupled from the band.

It follows that the boundary quantum phase transition
in this model
associated with the change in the ground state spin
is always of first order, although it will show
non-trivial finite-$T$ crossover behavior due to
residual ferromagnetic Kondo couplings.
Interestingly, this transition
does {\em not} affect the conduction electron phase shift,
as the phase shift is determined by
the two Kondo-screened spin-1/2 degrees of freedom only,
regardless of whether the two others form a decoupled singlet
or triplet.


{\it Applications of the double-impurity model}.
Two impurities coupled to a single orbital
may be hard to realize directly in metals,
however, the double-impurity model is obtained from the
usual two-impurity model \cite{2imp} in the limit of small
inter-impurity distance $R$.
In this limit, the screening channel associated
with the odd linear combination of conduction electrons
is weak,
and the physics of the double-impurity model can be
observed over a wide temperature range.
Experimentally, such a situation can be achieved with
magnetic ad-atoms on a metal surface manipulated by
scanning tunneling microscope (STM) techniques \cite{chen}.

Other applications can be constructed using
coupled or multi-level quantum dots \cite{vdw}.
In the former case,
each dot represents a single ``impurity'' spin,
and the inter-impurity interaction $I$ can be varied
either by tuning the inter-dot tunneling rate or
the direct exchange interaction.
For multi-level dots, we point out that the
single-channel situation considered here appears to be the
generic case for lateral devices \cite{qdotth4,vdw}
-- in contrast to a two-channel model appropriate for
vertical dots.
In fact, the results of Ref.~\onlinecite{vdw} indicate that
the KT transition discussed for the double-impurity model can
indeed be realized in a multilevel dot.




We thank A. Georges, F. Guinea, S. Kehrein, and K. Le Hur
for valuable discussions.
This research was supported by the DFG through SFB 484.



\begin{references}
\bibitem{nrgprob}
Data are shown for $N_s=650$ states per NRG step
and a discretization parameter $\Lambda=2$.
\bibitem{ferro} For both $J_{1,2}<0$ the Kondo couplings are irrelevant,
and consequently the model shows a first-order
transition between $S=0$ and $S=1$ ground states.
\bibitem{KTremark}
The Kosterlitz-Thouless transition separates two phases
where spin flips (here: of the second spin) are either relevant
(here: $I>I_c$) or irrelevant (here: $I<I_c$) in the
renormalization-group sense.
\end{references}
\end{document}